GIOVANNI BATTISTA RICCIOLI'S SEVENTY-SEVEN ARGUMENTS AGAINST THE MOTION OF THE EARTH: AN ENGLISH RENDITION OF *ALMAGESTUM NOVUM* PART II, BOOK 9, SECTION 4, CHAPTER 34, PAGES 472-7.

Christopher M. Graney

Jefferson Community & Technical College
1000 Community College Drive
Louisville KY 40272 (USA)
christopher. graney@kctcs.edu

In 1651 the Italian Jesuit Giovanni Battista Riccioli (1598-1671) published in his encyclopedic work on astronomy, the *Almagestum Novum*, 77 arguments against the Copernican movement of the Earth. These arguments are often mentioned in secondary sources, but a complete listing has not been readily available – thus one is provided here, in English. The 77 include interesting arguments from physics and astronomy that went on to become subjects of further investigation after the advent of Newtonian physics.

In 1651 the Italian Jesuit Giovanni Battista Riccioli (1598-1671) published in his encyclopedic work on astronomy, the *Almagestum Novum*, 77 arguments against the Copernican movement of the Earth. Indeed, a large portion of the book was dedicated to discussing arguments for or against different world system hypotheses, with Riccioli believing that the balance of argument favored a geo-heliocentric hypothesis such as that of Tycho Brahe (Figure 1). The 77 anti-Copernican arguments are a fascinating piece of the history of astronomy, physics, and science in general. The modern reader seeking to learn about the 77 will find that many authors mention them, but few provide details. Thanks to the advance of technology, a high-resolution copy of the *Almagestum Novum* is now widely available via the Internet[1]; thus the arguments are now available to all – in Latin. But the modern reader may not be inclined to dive into reading Latin material that has been characterized as being weak, religious rather than scientific in nature, and even "tedious or apparently stupid"[2], so this paper makes the 77 arguments available in English.

A number of authors mention Riccioli's 77 arguments briefly. At least one author provides a more detailed discussion of some of the 77.[3] The reader who searches the literature will find that various authors acknowledge that Riccioli

- ✔ was "a very learned man and good Astronomer"[4].

- ✔ was a "reknowned astronomer"[5].

---

1   G.B. Riccioli, *Almagestum Novum* (Bologna, 1651) <http://www.e-rara.ch/zut/content/pageview/140188>.

2   Bruce Eastwood, review of Edward Grant's "In defense of the Earth's centrality and immobility: Scholastic reaction to Copernicanism in the seventeenth century", *Isis*, 76 (1985), 378-9.

3   Edward Grant, "In Defense of the Earth's Centrality and Immobility: Scholastic Reaction to Copernicanism in the Seventeenth Century", *Transactions of the American Philosophical Society*, New Series, Vol. 74, No. 4 (1984), 12.

4   Robert Hooke, *An Attempt to Prove the Motion of the Earth from Observations* (London, 1674), 5.

5   J. L. E. Dreyer, *History Of The Planetary Systems From Thales To Kepler* (Cambridge University Press: Cambridge, 1906), 419.

- ✔ was a true scientist, a scientist who undertook great efforts in order to obtain precise data[6], at a time when many of his contemporaries were natural philosophers in the medieval sense[7].

- ✔ produced, in his *Almagestum Novum*, the lengthiest, most penetrating, and authoritative analysis of the question of Earth's mobility or immobility made by any author of the sixteenth and seventeenth centuries.[8]

- ✔ claimed to be seeking only the Truth, and to be unprejudiced by any authority, and that in this regard "his words ought to be taken at face value".[9]

However, they characterize the 77 arguments as

- ✘ being overly earnest and zealous.[10]

- ✘ being a "sterile exercise".[11]

- ✘ motivated by religious faith more than scientific argument[12], so that biblical and theological arguments were decisive[13].

- ✘ being based on Aristotelian classification of motions, and Aristotelian concepts such as gravity and levity[14].

---

6  J. L. Heilbron, *The Sun in the Church: Cathedrals as Solar Observatories* (Harvard University Press: Cambridge Massachusetts, 1999), 180-81.

7  "In Defense of the Earth's Centrality and Immobility...", 12.

8  Edward Grant, *Planets, Stars, and Orbs: The Medieval Cosmos, 1200-1687* (1996), 652.

9  Alfredo Dinis, "Giovanni Battista Riccioli and the Science of His Time", in *Jesuit Science and the Republic of Letters*, ed. by Mordechai Feingold (MIT Press: Cambridge, Massachusetts, 2003), 199.

10  *An Attempt to Prove the Motion of the Earth...*, 5.

11  Albert van Helden, "Galileo, Telescopic Astronomy, and the Copernican System", in *The General History of Astronomy*, ed. M. A. Hoskin, 4 vols., 2A (Cambridge University Press, 1984), 103.

12  *The Sun in the Church...*, 184. Also Christopher M. Linton, *From Eudoxus to Einstein: A History of Mathematical Astronomy* (Cambridge University Press, 2004), 226-227.

13  *Planets, Stars, and Orbs...*, 63.

14  Alfredo Dinis, "Was Riccioli a Secret Copernican?", in *Giambattista Riccioli e il Merito Scientifico dei Gesuiti Nell'eta Barocca*, M. P. Borgato (Firenze: Leo S. Olschki, 2002), 63.

> ✗ lacking in any real argument for the geocentric system beyond simply the Bible and the authority of the Roman Catholic Church.[15]

Indeed, writers sometimes describe Riccioli as being a secret Copernican, and thus the 77 anti-Copernican arguments as being a formality.[16]

Even though many authors mention the 77, a complete listing is exceedingly difficult to find. Thus what is provided here is an English rendition of *Almagestum Novum* Part II, Book 9, Section 4, Chapter 34, pages 472-7. It is not of a close translation of Riccioli's arguments (and the Copernican answers to them), but a tabular listing of them in the briefest form possible. The goal here is to make the arguments known, and to provide scientific analysis of some, in the hopes of generating interest in them. Occasionally, when Riccioli is succinct, a close and brief translation is useful; quotation marks indicate these. Chapter 34 is Riccioli's synopsis (still lengthier than what is presented here) of the arguments; each argument is numbered in the same manner as it is numbered here; Riccioli provides marginal notes for each, directing the reader who enjoys Latin to the places in the *Almagestum Novum* where the reader will find more detailed treatments.

There are notable discrepancies between the 77 and what has been written about them that will be apparent to the reader. Perhaps most prominent is that only two are at all religious in nature, and these are both minor. In contrast, a religious argument – an appeal to the power of God – features importantly in the Copernican response to some arguments.

The arguments range from arguments based on physical experiments with falling bodies, to arguments based on telescopic observations of the stars, to arguments based on simplicity of motion, to arguments based on the wind. Some are simplistic, easily refuted by common motion, common sense, or basic astronomical knowledge. Some are challenging, and can only be answered by extensive experimentation.

Riccioli does not present the arguments as being all of equal weight. He labels only a handful as being key arguments to which the Copernican hypothesis has no good answer. These include:

---

15    "Giovanni Battista Riccioli and the Science...", 209.

16    *History Of The Planetary Systems...* , 419; "Was Riccioli a Secret Copernican?"

- arguments based on simplicity, proportion, and economy of motion.

- arguments based on the effect of a rotating frame of reference on the movement of artillery projectiles and falling bodies.

- arguments based on observations of the stars (including telescopic observations) – these being the ones that Riccioli says prompt Copernicans to appeal to Divine Omnipotence.

The Copernicans had their own simplicity arguments, and determining whose idea of simplicity was the more true was impossible. As Robert Hooke would comment a couple of decades after the *Almagestum Novum*:

> What way of demonstration have we that the frame and constitution of the World is so harmonious according to our notion of its harmony, as we suppose? Is there not a possibility that things may be otherwise? nay, is there not something of a probability?[17]

But Riccioli's other two key arguments were not easily answered. The effect of Earth being a rotating frame of reference (the Coriolis force) was not observed until the 19th century, despite various attempts to do so since the 17th century.[18] Nor did Astronomers obtain a full understanding of the nature of telescopic star images until the 19th century.[19] Indeed, Owen Gingerich has argued that what brought about the acceptance of the Copernican hypothesis was not observational "proofs" such as the telescopic discovery of the phases of Venus or the moons of Jupiter; these could be incorporated into the Tychonic geocentric hypothesis easily enough. Rather, it was Newton's development of a coherent theoretical framework that explained the Copernican hypothesis, but not the Tychonic one, that persuaded astronomers that the Copernican hypothesis was correct, even in the *absence* of observational proofs. Gingerich notes that scientists did not dance in the streets and hold grand celebrations in 1838 when Bessel measured annual parallax, or in 1851 when Foucault's pendulum clearly demonstrated that Earth was a rotating frame of reference – the matter had already been settled by Newton.[20] Thus Riccioli's most powerful arguments among the 77 were essentially unanswerable in their time, and would live on, after Newton and the broad acceptance of the Copernican hypothesis, to become matters of further scientific investigation.

---

17  *An Attempt to Prove the Motion of the Earth...*, 3.

18  See note 28.

19  Christopher M. Graney and Timothy P. Grayson, "On the Telescopic Disks of Stars: A Review and Analysis of Stellar Observations from the Early Seventeenth through the Middle Nineteenth Centuries", *Annals of Science*, iFirst 27 October 2010, DOI: 10.1080/00033790.2010.507472 (print version in press).

*The Seventy-seven Anti-Copernican Arguments from the* **Almagestum Novum** *with Copernican answers to them, in brief, with comments.*

|  | ARGUMENT | COPERNICAN ANSWER |
|---|---|---|
| #1 | *The rate of increase in speed of falling heavy bodies, as determined by experiment, is incompatible with the hypothesis that all natural motion is circular* [henceforth "NMC hypothesis"]*, the only viable hypothesis that could provide a theoretical explanation for the diurnal motion.* [this is Riccioli's "physico-mathematical" argument.] | *No solid Copernican answer against this argument.* |
|  | Comment: In his *Dialogue Concerning the two Chief World Systems*, Galileo proposed that the apparent linear acceleration of a stone falling from a tower might be the result of two uniform circular motions – the diurnal rotation of Earth, and a second uniform circular motion belonging to the stone (with the same circumferential speed as the diurnal motion at the top of the tower, but centered on a point located half-way between the Earth's center and the top of the tower).[21] Thus, Galileo says, | |

---

20  Owen Gingerich, *God's Universe* (Harvard University Press: Cambridge Massachusetts, 2006), 94.

21  Galileo Galilei, *Dialogue Concerning the Two Chief World Systems: Ptolemaic and Copernican,* translated and with revised notes by Stillman Drake, foreword by Albert Einstein, introduction by J. L. Heilbron (Modern Library/Random House: New York, 2001), 189-94.

| ARGUMENT | COPERNICAN ANSWER |
|---|---|
| | [T]he true and real motion of the stone is never accelerated at all, but is always equable and uniform.... So we need not look for any other causes of acceleration or any other motions, for the moving body, whether remaining on the tower or falling, moves always in the same manner; that is, circularly, with the same rapidity, and with the same uniformity....[22] <br><br>Galileo goes on to say that the movement of a falling body is either exactly this, or very near to it;[23] and that – <br><br>straight motion goes entirely out the window and nature never makes any use of it at all.[24] <br><br>Thus here is a physics hypothesis to explain motion in the Copernican theory – all natural motion, including that of heavy objects such as stones, is circular; the motion of Earth is thus natural; natural straight-line motion does not exist, and what appears to be such motion, like the falling stone, is the result of a combination of circular motions. <br><br>However, a rigorous analysis of Galileo's NMC hypothesis leads to experimentally testable predictions regarding the rate of acceleration of a falling body. Riccioli (and his team of Jesuits: Grimaldi and others) devised precise experiments to measure this rate of acceleration. The rate they determined experimentally (9.6 m/s$^2$, differing from the accepted modern value of 9.8 m/s$^2$ by a mere 2%) disagreed strongly with the rate predicted by the NMC hypothesis.[25] <br><br>Riccioli's first few arguments are directed against the NMC hypothesis; the Copernican theory is weaker in the absence of a theoretical framework to explain its motions.[26] |

---

22   *Dialogue*, 193.

23   *Dialogue*, 194.

24   *Dialogue*, 194.

25   *The Sun in the Church...*, 178-81.

26   Owen Gingerich has emphasized in his writings that credible scientific explanations hang together in a tapestry of coherency that supports observations: "Truth in Science: Proof, Persuasion, and the

|    | ARGUMENT | COPERNICAN ANSWER |
|----|----------|-------------------|
| #2 | *The same as argument #1, but including the issue of annual motion against the NMC hypothesis as well.* | *No Copernican answer which is not sophistical, and full of foolish evasion.* |
| #3 | *If Earth had a diurnal rotation, heavy bodies falling near the equator would have a fundamentally different motion than identical bodies falling near the poles under identical conditions.* | *Three possible Copernican answers – all rejected.* [Two do not merit discussion (these having to do with magnetism and air).] *The third is that a heavy body moves with two motions: a downward motion owing to the body's gravity, and a circular "common motion".* [But, says Riccioli, this answer is contrary to the essence of the NMC hypothesis.] |
| #4 | *The same as argument #3, but including the annual motion, which complicates even the comparatively simple case of a body falling at the poles of a diurnally rotating Earth.* | |
| #5 | *The same arguments as #1 through #4, but applied to light bodies whose natural motion is upwards.* | |
| #6 | *Heavy bodies naturally fall to Earth along a line that is straight and perpendicular to ground. If launched perpendicularly upwards, they fall back upon the location from which they were launched. If the Earth had diurnal and annual motions,* | *The Copernican answer is weak, being that falling objects only appear to move linearly.* |

---

Galileo Affair", *Perspectives on Science and Christian Faith*, 55 (June 2003), 85-86; *God's Universe* (Harvard University Press: Cambridge Massachusetts, 2006), 91-95.

| ARGUMENT | COPERNICAN ANSWER |
|---|---|
| *these bodies would follow curved trajectories.* | |

Comment: Argument #6 is the first of several "Coriolis force" arguments. The Coriolis force is an illusory "force" that exists because the surface of the Earth is a rotating, spherical frame of reference, not a translating, flat frame of reference. It is thought to cause the rotation seen in large weather patterns. Long-range artillery is deflected because of it.[27] And, it causes falling bodies to follow a slightly curved path. However, the effect is much more difficult to detect than calculations would suggest.[28]

In discussing the Copernican answer to this argument, Riccioli insists that it is *physical*

---

[27] Jerry B. Marion, *Classical Dynamics of Particles and Systems* (Academic Press/ Harcourt Brace Jovanovich: Orlando, Florida, 1970), 343-56.

[28] One might calculate the expected easterly deviation of a body falling from a tower at the equator as follows: If the tower has height $h$, and the Earth has radius $R$ and rotates in time $T$, then the speed of the bottom of the tower is $v = 2\pi R/T$, while the top of the tower exceeds this speed by a fraction equal to $h/R$. Thus the top of the tower exceeds the bottom by speed $s = (2\pi R/T)(h/R) = 2\pi h/T$. The time required for a heavy body to fall from the tower top is $t = (2h/g)^{1/2}$ (this could also simply be measured ). Thus the easterly deflection is $d = st = (2\pi/T)(2h^3/g)^{1/2}$. Thus an object falling from a 100 m tall tower (i.e. the Torre degli Asinelli in Bologna, used by Riccioli for experiments involving falling bodies) at the equator should be deflected eastward by approximately 3.3 cm. This would drop to zero at the poles. (A proper modern mechanics treatment of the "Coriolis force" tells us that the easterly deflection is actually $d \approx 2/3(2\pi/T)cos(\lambda)(2h^3/g)^{1/2}$, where $\lambda$ is the latitude. This yields a 2.2 cm deflection at the equator, dropping to 1.6 cm at 45° N. Latitude (Bologna), and to zero at the poles (*Classical Dynamics...*, 350).)

Thus the deflection effect would be expected to be small, but would not seem to scientists of the time to be immeasurable (Newton said the effect would be "very small, and yet I am apt to think it may be enough to determine the matter of the fact"), since issues with the influence of the air should be avoidable by dropping very dense, heavy bodies. Robert Hooke attempted to detect the deflection in 1680, and various other less-than-successful attempts followed. In 1831 F. Reich dropped objects through a distance well over 100 m in an enclosed mine shaft in Freiberg, Germany, and recorded a definite deflection. Experiments of this nature were still being performed in the early 20th century, for there are apparently subtle effects that make detecting the deflection very difficult. (See *Classical Dynamics...*, 350; William F. Rigge, "Experimental Proofs of the Earth's Rotation", *Popular Astronomy*, 21 (1913), 208-212; Edwin H. Hall, "Do Falling Bodies Move South", *Physical Review*, 17 (1903), 179-190; Walter William Rouse Ball, *An Essay on Newton's 'Principia'* (MacMillan and Co.: New York, 1893), 139-153, which contains much on Newton and Hooke, including the Newton quote above, 143.) Hall notes how "curious things may well occur [p. 188]" in these sorts of experiments. Thus a relatively simple experimental test that a scientist of Riccioli's era would expect to reveal Earth's motion, generally fails to do so.

| ARGUMENT | COPERNICAN ANSWER |
|---|---|
| | *evidence* that must be the deciding factor. If such evidence cannot be relied upon, then "all physical knowledge will be destroyed".[29] |
| #7 *A moving Earth means less economy of motion: bodies would not follow the shortest routes when returning to their natural places, as their routes would be curved rather than linear.* | *The Copernican answer is to deny the necessity of following the shortest route.[30] (Lack of economy of motion is a general problem afflicting the Copernican theory; it multiplies overall motions in the universe.)* |
| #8 *A moving Earth invalidates the standard explanation for the downward motion of heavy bodies – that they tend, through the shortest route, to the place they ought to occupy in the system of elements. No comparably excellent explanation for such motion exists if the Earth moves.* | *No sufficiently strong Copernican answer against this argument.* |
| Comment: Riccioli here considers and rejects the explanation that objects fall downward owing to attraction between matter and matter, noting that a stone dropped down a well is not attracted to the walls of the well. In later arguments, Riccioli provides a Copernican answer to this type of argument – that being that the system of elements moves with the Earth. See argument #39 and following. ||
| #9 *The movement of the Earth requires more types of motion.* | *No sufficiently strong Copernican answer against this argument. Copernicans* |

---

29     "tota scientia Physica peribit"

30     Riccioli apparently could not resist adding a little gratuitous commentary here: "who does not see that this answer has been raked up from the muck, not owing to real insight into the nature of heavy bodies, but simply to protect the hypothesis of a moving Earth? [at quis non videt id mendicatim conquisitum non ex natura Grauium, sed ad tuendam hypothesim motus terræ?]"

| ARGUMENT | COPERNICAN ANSWER |
|---|---|
| *"[M]ore movements are imposed on the system of the universe, if Earth be moved, than if it rests...."[31]* | *attempt to argue that if the Earth is not moved then the daily motion is multiplied in the fixed stars and in the planets.* |
| Comment: Riccioli counters the Copernican answer by stating that all motions in the heavens are of one kind, from east to west; apparent easterly motion is owed to simply slower westerly motion. | |
| #10 *Imagine a great weight, dropped from on high, paying out a chain as it falls. If the Copernican hypothesis is correct, the chain would not be extended straight down to Earth, but would be curved to the east.* | |
| Comment: This is another version of the Coriolis Force argument (see Argument #6). Riccioli acknowledges that this contrived argument (more of a thought experiment than anything else – an angel would have to perform the experiment) is of limited value. | |
| #11 *If Earth moves, then no straight lines that we may construct can be known to be truly straight, "But in the presence of God and the angels they might be different shapes etc."[32]* | |
| #12 *If Earth moves, then the clouds and the birds in the air would be seen to fly west, as they were left behind by the Earth.* | *The Copernicans answer that any body composed of the elements earth and water, before its private motion (if it has any such), has also motion common to the whole earth and water, by which equal velocity, or through like arcs, carries the* |

---

31 "quia reuera plures motus ponuntur in Mundi systemate, si Tellus moueatur, quam si quiescat"

32 "Sed coram Deo & Angelis essent diuersę figuræ &c."

| ARGUMENT | COPERNICAN ANSWER |
|---|---|
|  | *whole into the East. This may not be seen by us, because that motion is likewise common to us.* |
| Comment: Argument #12 is the first argument against the Earth's motion for which Riccioli states that the Copernicans have a good answer – that being the "common motion". Riccioli is listing all arguments against Earth's motion – not just arguments he thinks are valid. Riccioli will go on to list a number of arguments which are easily refuted by "common motion" (not to mention common sense). ||
| #13 *If Earth moves, then it should be more difficult to move towards the east than towards the west, owing to air resistance...* | *The Copernicans answer that common motion applies to air, too.* |
| #14 *...and there should be a continuous wind from the west...* | *Common motion applies to air.* |
| #15 *...and there should be various other effects caused by that motion...* | *...all of which can be dismissed with the answer of common motion.* |
| #16 *If Earth rotates, a cannon ball launched toward the west should travel further than an identical shot to the east, for the cannon pursues the eastern ball and recedes from the western one. But this is contrary to the experiments of Tycho and Landsgrave.* |  |
| Comment: Riccioli discusses the answer to this argument, which he states in terms of the motive force added to or subtracted from the ball, etc. but which essentially is a variation on the common motion idea. ||
| #17 *A cannon ball launched in the direction of the plane of the meridian (due north or* | *No solid Copernican answer against this argument.* |

| ARGUMENT | COPERNICAN ANSWER |
|---|---|
| *south) will have a different trajectory if the cannon is nearer the poles than if it is nearer the equator, owing to the slower speed of the ground near the poles. But this is contrary to the experiments of Tycho.* | |
| Comment: Another Coriolis Force argument (see Arguments #6 and #10). Riccioli adds that the only answer to this argument is that perhaps such an experiment has never been properly performed (apparently Tycho's experiments were not completely convincing). However, he says, the experiment is possible – the effect should not be insensible if the motions involved are sufficiently violent (that is, for artillery of sufficient range). | |
| #18 *If Earth rotates, the ball from a cannon aimed at a western target will hit below the mark, while the ball from a cannon aimed at an eastern target will hit above the mark. But this is contrary to experience.* | *Galileo has answered this argument, calling such experiments into doubt.* |
| Comment: At first glance this Argument appears to be a variation on #16, but it is much different. #16 deals with motion towards the east or west, as though the surface of the Earth moved linearly at a fixed rate (that is, with translational motion). This Argument deals with direction changes owing to Earth's rotational motion – the line from a cannon's muzzle to a target changes as Earth turns, while the flying ball's trajectory does not, with the results being as Riccioli states. As this Argument is based on Earth being a rotating frame of reference, it has more in common with the Coriolis Force Arguments seen so far (#6, #10, # 17) than the Common Motion Arguments (#12 through #16). | |
| Galileo addresses this question in his *Dialogue*, arguing that the effect would be about one inch of deviation at a range of 500 yards – too small to measure, a cannon being accurate to no better that a yard at that range.[33] But, Riccioli notes, movement of the | |

---

33  *Dialogue...*, p. 209-212.

| | ARGUMENT | COPERNICAN ANSWER |
|---|---|---|
| | Earth should conceivably be detectable by this sort of experiment. | |
| #19 | *If Earth rotates, the range of a cannon ball will be less if launched toward the pole of the world than if launched east or west. But this is contrary to experience.* | *There is no Copernican answer that weakens this argument.* |
| | Comment: Another Coriolis Force Argument. Riccioli cites Grimaldi for his work on the physics of this Argument and refers the reader to elsewhere in the *Almagestum Novum* for details. | |
| #20 | *If Earth rotates, the range of a cannon ball will be less if launched towards west than towards the east. But this is contrary to our experiments.* | *No solid Copernican answer against this argument.* |
| | Comment: A variation on #18, for if a projectile will hit above the mark to the east, and below it to the west, it should travel farther to the east than to the west. Thus, another Coriolis Force Argument. | |
| #21 | *If Earth rotates, a thing could move simultaneously in two directions – something moving to the west also moves into the east owing to motion with Earth. But this is impossible.* | *First, nothing can have double motion in that it cannot simultaneously approach and recede from the same fixed point in the universe. In the case of Earth, something moving to the west simply moves east less swiftly. Second, this same argument can be tossed back to the geocentrists, who have no difficulty with this issue in regards to the motions of the heavens.* |

|     | ARGUMENT | COPERNICAN ANSWER |
| --- | --- | --- |
| #22 | *A moving Earth multiplies motions, for every object on Earth has a motion as part of the common motion.* | *A fixed Earth multiplies motions in the heavens.* |
|     | Comment: Riccioli notes that fewer motions are required if it is the heavens that move. Presumably he believes there to be fewer stars in heaven than grains of sand on Earth. #22 and #23 are both "economy of motions" arguments that seem very similar. ||
| #23 | *There is less multiplication of the real movements if daily motion is attributed to the stars, and annual motion to the Sun, than if these are attributed to Earth.* | *No solid Copernican answer against this argument.* |
| #24 | *If Earth moves, then motions which are manifestly apparent to us are, without necessary reason, destroyed and replaced with movements which are not apparent. This is certainly absurd.* | |
| #25 | *If Earth moves, then more variation of motion is attributed to a single moving thing than if the stars are what moves.* | *No firm Copernican answer to this argument.* |

|     | ARGUMENT | COPERNICAN ANSWER |
| --- | --- | --- |
| #26 | *The Earth is most dense as well as most heavy, and so most resistant to motion.* | *Weight does not resist circular motion.*[34] |
| #27 | *The speed of the Earth's rotation is so great it might overwhelm the flight of birds, the movement of ships, etc.* | *Common motion.* |
| | Comment: Riccioli does not reject the Copernican answers to #26 and #27, but he does include comments about just how heavy is Earth and just how great are the speeds associated with Earth's motion. | |
| #28 | *If Earth moves, then we should experience a continuous wind toward the west.* | *Common motion applies to air; and, there are such winds in the tropics.* |
| #29 | *If Earth moves, then buildings could not stand and objects not anchored to Earth should fly off.* | |
| #30 | *If Earth moves, then we should feel the motion within ourselves.* | |
| #31 | *If Earth turns into the east, eastern mountains should descend, and western ones ascend.* | |

---

34   This suggests the ideas of Jean Buridan:

> ...God, when He created the world, moved each of the celestial orbs as he pleased, and in moving them He impressed in them impetuses which moved them without his having to move them any more except by the method of general influence whereby he concurs as a co-agent in all things which take place; "for thus on the seventh day He rested from all work which He had executed by committing to others the actions and the passions in turn." And these impetuses which He impressed in the celestial bodies were not decreased nor corrupted afterwards, because there was no inclination of the celestial bodies for other movements. Nor was there resistance which would be corruptive or repressive of that impetus.

Buridan, "The Impetus Theory of Projectile Motion", translated from Latin into English by Marshall Clagett, in *A Source Book in Medieval Science* by Edward Grant, editor (Cambridge Massachusetts: Harvard University Press, 1974), 277.

| | ARGUMENT | COPERNICAN ANSWER |
|---|---|---|
| #32 | *If Earth rotates, then a star viewed from the bottom of a well should pass out of view in the blink of an eye, owing to the rapidity of Earth's motion.* | |
| #33 | *If Earth rotates, gnomons built on the Tropic should cast shadows at noon on the Summer solstice, which they do not.* | |
| | Comment: Riccioli says arguments #29 - #33 are mathematically incorrect and refers the reader to elsewhere in the *Almagestum Novum* for details. | |
| #34 | *"The eclipse of the sun at the death of CHRIST was total for three hours: but if Earth by daily motion might have been turned, it might not have remained total for three hours, in fact the rotation of the Earth might have immediately carried away Palestine into another position, from which the Sun might have been able to be seen. Therefore."*[35] | *The moon could move so as to compensate for Earth's rotation.* |
| | Comment: Argument #34 is one of few that Riccioli lists that relate to Christian scripture or religious matters. | |
| #35 | *Circular motion is unnatural for earthly objects, so it is unnatural for the whole Earth as well.* | *Circular motion is indeed natural for earthly objects, as they all move in circular paths, and only appear to move straight to us who are moving with the* |

---

35    "Eclipsis Solis in morte CHRISTI fuit totalis per tres horas: sed si Tellus diurno motu conuersa fuisset, non durasset totalis per tres horas, Telluris enim vertigo subtraxisset statim Palæstinam in situm alium, ex quo Solem videre potuisset. Ergo."

| | ARGUMENT | COPERNICAN ANSWER |
|---|---|---|
| | | *Earth. Moreover, Aristotle allowed that fire might have perpetual circular motion, even if it was not natural.* |
| #36 | *A moving Earth removes from the Universe the simple movement of the things up and down.* | *This is not true:* apparent *movement up and down remains.* |
| #37 | *What starts the Earth's motion?* | *The motion is intrinsic and natural.* |
| #38 | *A moving Earth renders unnatural the motions of heavy and light bodies, while rendering circular motion natural.* | *The Copernicans deny these definitions of natural motion.* |
| #39 | *According to Aristotle, heavy bodies tend toward, and light bodies recede from, the center of the universe, not the center of the Earth.* | *Heavy bodies carried by the Earth tend towards the center of the Earth – the center of the heaviest body. Light bodies tend toward the circumference of the elemental system, which Aristotle has not proven to be concentric to the universe.* |
| | Comment: Riccioli mentions both Galileo and Kepler in connection with this response. He says Galileo's response[36] is not bad, but criticizes Kepler. The idea that the Earth lies at the center of a spherical elemental system that circles the sun as a whole, and within which the Aristotelian elements and physics is valid (Figure 2), plays a prominent role in the Copernican answers to a number of the upcoming arguments. | |
| #40 | *Light bodies ascend along a line that is perpendicular to both Earth's surface and the sphere of the highest heaven. Thus they ascend from the center of the Earth and the* | *Light bodies ascend not towards the sphere of the highest heaven, but toward the sphere of the elemental system, which Copernicans contend may not be* |

---

36  See, for example, *Dialogue*, 285.

|     | ARGUMENT | COPERNICAN ANSWER |
| --- | --- | --- |
|     | *center of the Universe.* | *concentric with the Universe.* |
| *#41* | *Weight and levity are not given to bodies so that they may be united to things like themselves, but so that they may retain or regain their determined place in the universe. For heavy bodies this is in the center of the Universe; for light bodies this is around the center of the Universe. They do not have these places if Earth has an annual motion.* | *The places of heavy and light bodies are not determined within the Universe, but within the elemental system.* |
| *#42* | *The Earth must be the center of the Universe, for there is no explanation as to what would keep it in any other position.* | *The entire Earth has a natural circular motion about the center of the Universe. Kepler says that the Earth as a whole is not heavy.* |
| *#43* | *If Earth were shifted towards the moon, heavy bodies would still tend toward the center of the Universe, not towards the Earth.* | *Aristotle has not shown this.* |
| *#44* | *The lowest place belongs to the heaviest and lowest of bodies. The Earth is the heaviest body. The center of the Universe is the lowest place. Thus Earth lies at the center of the Universe.* | *It is the lowest place in the elemental system, not the absolute lowest place, that belongs to heavy bodies.* |
| *#45* | *Heavy bodies are those that tend toward the center of the universe, and light bodies those that tend away from the center. These definitions are ruined by an annually* | *Heavy bodies are those that tend toward the center of the elemental system, and light bodies are those that tend away from* |

| | ARGUMENT | COPERNICAN ANSWER |
|---|---|---|
| | *moving Earth.* | *it.* |
| *#46* | *Unless the center of the Earth and the elemental system is the center of the Universe, the positive Levity of light bodies is reduced to simply lack of Weight.* | *The physical place of light bodies, the place of elemental fire, is the space between the Moon and heavier elements, regardless of where the whole elemental sphere is placed.* |
| *#47* | *If Earth is not at the center of the universe, then a heavy body descending to the center of Earth could be receding from the center of the universe, and vice versa for a light body. This confounds the definitions of Heavy and Light.* | *The definitions supposedly confounded apply only in the traditional Aristotelian system.* |
| *#48* | *Weight and Levity is attached to bodies, in terms of the place to which they tend, at which place they rest. "But they might never rest if Earth with the elements rolls through the annual orb."*[37] | *Weight and Levity is attached to bodies, in terms of which stands over or under the other in the elemental system.* |
| *#49* | *In the Copernican hypothesis, centers and the positions of the centers are unnecessarily multiplied, as one is the center of the Universe, and a different one is the center of the Earth and elemental system.* | *Two Copernican answers to this: first, there is no a priori reason for only one center; second, and more forcefully, the geocentric hypothesis has two centers – for while the Earth is the center of the Universe, the sun is the center of the planetary system.* |
| | Comment: The Copernican answer is is in reference to the Tychonic hypothesis, in which the planets circle the sun while the sun circles the Earth. This was the only sort of | |

---

37     "At nunquam quiescerent si Tellus cum elementis volueretur per orbem annuum."

| | ARGUMENT | COPERNICAN ANSWER |
|---|---|---|
| | geocentric hypothesis consistent with telescopic observations, which showed that planets such as Venus circle the sun. | |
| #50 | *"All men observing the heaven from any vantage point of the Earth, consider the heaven to be up, and Earth down; But this judgment is false, if Earth is outside of the center of the Universe."*[38] | *As determined by physics and the senses Earth is the center, and up and down remain; but Earth is not the center overall, as determined by mathematics.* |
| #51 | *The Earth is lowest, not only of the elements, but of all the Universe's bodies. Therefore, it must be in the lowest place, not only in the elemental system, but in the Universe. And that place is the center of the Universe.* | *The Earth is not the lowest of all the Universe's bodies, for it contains men and other living things.* |
| #52 | *The Copernican hypothesis gives excessive license to place Earth anywhere.* | *Any place that saves the phenomena is a proper place for Earth.* |
| #53 | *If Earth is not the center of the Universe, then Hell is not at the lowest place, and someone going to Hell could conceivably ascend in doing so.* | *Hell is a place defined by comparison, to this world on which men travel and God's Heaven. The relationship between Heaven, Hell, and the world of men is not affected by whether Earth moves.* |
| #54 | *"If Earth is in the Annual Orb with the elements, the order of the system of Planets and elements is perverted..."*[39]; *the sun and moon cease to be planets, there are six* | *This argument is relevant only for those who value the order of the things according to archetypical reckoning.* |

---

38   "Omnes ex quauis Terrę parte cælum spectantes, æstimant cælum esse sursum, et Terram deorsum; At hoc iudicium falsum esset, si Tellus esset extra centrum Mundi."

39   "Si Tellus sit in Orbe Annuo cum elementis, peruertitur ordo systematis Planetarij et elementaris...."

|   | ARGUMENT | COPERNICAN ANSWER |
|---|---|---|
|   | *planets rather than seven, etc.* |   |
| #55 | *All the heavenly phenomena are saved by supposing Earth to be in the center of the Universe.* | *All the heavenly phenomena are saved by supposing Earth circles the sun annually and rotates daily.* |
| #56 | *"It is necessary to attribute more motions to Earth, with more changes in the stars etc."*[40] | *The Copernicans accept that the Earth has more motions, but reject that this implies changes in the stars* [such as annual parallax]. |
| #57 | *If Earth did not lie at the center of the heavens, observers on Earth might not see a complete hemisphere of heaven .* |   |
| #58 | *The fixed stars towards which Earth moves should grow larger.* |   |
| Comment: The Copernican answer to #57 and #58 is for the stars to be so distant that the size of the Earth's orbit is negligible by comparison. |||
| #59 | *"The eastern gnomon shadows at equal height of the Sun from the horizon might not be equal to the western ones."*[41] | *No, as both are equally distant from the Sun.* |
| #60 | *The changes of the days and of the nights would not happen as they do.* | *This idea is wrong and simply a result of ignorance of the Copernican hypothesis.* |
| #61 | *Eclipses of the Moon might not always happen with the Moon opposite the Sun in the Zodiac.* | *In the Copernican hypothesis, in an eclipse the Earth is still always interposed between the Moon and Sun on a line.* |

---

40  "Oporteret plures motus Terræ attribuere, cum magis mutationibus in stellis &c."

41  "Vmbræ Orientales gnomonum in pari altitudine Solis ab horizonte non essent æquales occidentalibus."

| | ARGUMENT | COPERNICAN ANSWER |
|---|---|---|
| #62 | *Eclipses of the Moon might not be equally visible from opposite horizons* [where the sun is setting/rising]. | |
| | Comment: Here Riccioli refers the reader to the answer to #61, and remarks on the ignorance of anyone who would advance this argument. Arguments #59-#61 seem to be primarily arguments from ignorance. | |
| #63 | *In the Copernican hypothesis, the Earth completes nearly 365¼ daily rotations in one annual revolution about the sun. This disjunction between these two rates is too high according to physics. The daily rotation should be slower.* | *The disjunction is a matter of mathematics more than physics.* |
| #64 | *If Earth be moved through the Annual orb, then a sensible difference should be detected in the altitude of Fixed stars over 6 months – a notable parallax – at least in stars nearer to the* [ecliptic] *pole. But the astronomers have observed no parallax in the Fixed stars.* | |
| #65 | *A parallax in Sirius should be detectable between the equinoxes.* | |
| #66 | *"But surely conspicuous parallax might be seen in the apparent size of the Fixed stars."*[42] | |
| | Comment: For each of the above three arguments, Riccioli notes that the Copernican | |

---

[42] "At certe insignis parallaxis sentiretur in magnitudine apparenti Fixarum."

| ARGUMENT | COPERNICAN ANSWER |
|---|---|
| answer is that these effects will vanish if the stars are sufficiently distant. | |
| #67 *The annual motion of the Earth requires that, for there to be no sensible parallax, the fixed stars be a huge distance from the Earth and the center of the Universe. Thus the globe of the stars will be immense beyond credibility...* | |
| #68 *...and thus between Saturn and the fixed stars will be immeasurable space, idle and unoccupied...* | |
| #69 *....and thus the sun will be too distant from the stars to illuminate them...* | |
| #70 *...and thus the sizes of the fixed stars will be beyond credibility – comparable to the size of the Annual Orb.* | |
| Comments: Riccioli states that the Copernican answer to the issues of the immensity of the sphere of the stars and of the stars themselves is that the immensity is not incredible, but admirable: "[I]t may more greatly point out Divine Omnipotence and Magnificence."[43] Riccioli stops short of calling this sort of answer invalid, but he still criticizes the Copernican's use of it – his opinion is that it is a falsehood that cannot be completely refuted, yet cannot satisfy the more prudent man. He remarks upon the Copernicans resorting to "improbable subtleties"[44] in defending the space between | |

---

43   "et Diuinam Omnipotentiam ac Magnificentiam magis commendet" (quote found under argument 67). For discussion of a Copernican, Philips Lansbergen, who advocated such a view, see Reink Vermij, "Putting the Earth in Heaven: Philips Lansbergen, the Early Dutch Copernicans and the Mechanization of the World Picture", in *Mechanics and Cosmology in the Medieval and Early Modern Period*, M. Bucciantini, M. Camerota, S. Roux, editors (Florence, 2007).

44   "improbabilibus subtilitatibus" (Argument LXVIII, *Responsiones*)

| ARGUMENT | COPERNICAN ANSWER |
|---|---|
| Saturn and the stars while being quite willing to define what God and Nature would choose to do in other situations. He notes that the Copernicans deny that the sun illuminates the stars.<br><br>The issue of the physical sizes of the stars arises from the appearance of stars through small-aperture telescopes: seen through such telescopes, stars appear as disks of measurable size (Figure 3), and therefore the more distant they are, the larger they must be. If the stars are so distant that parallax is insensible, they must be immense. Riccioli included in the *Almagestum Novum* tables of and discussion about the apparent diameters of stars, measured telescopically, and of their calculated physical sizes at the distances required by the Copernican hypothesis, showing their immensities.[45]<br><br>Riccioli notes that the Copernicans respond that the heliocentric immense size of the stars is no more incredible than the geocentric great speed of them; but Riccioli says this is demonstrably incorrect.[46] Finally, Riccioli declares that if God's purpose with the stars is to make Himself known to us, he might choose to do that in a manner which is apparent (i.e. their great visible speed in the geocentric hypothesis), rather than in a manner that hides the stars' vast sizes behind such a small appearance. | |
| #71 *Sensible refraction is observed in the fixed stars – at least 30' at the horizon. If Earth revolves in the annual orb, the distance of the fixed stars is so great that no sensible refractions of them should occur, on account of the inclination of the incidental rays into our air...* | |

---

45  Christopher M. Graney, "The Telescope Against Copernicus: Star Observations by Riccioli Supporting a Geocentric Universe", *Journal for the History of Astronomy*, 41 (2010), 457-461.

46  A valid response, as Riccioli calculated that a single star in the heliocentric hypothesis could conceivably exceed the size of the entire universe as determined by Tycho. "The Telescope Against Copernicus...", 460-461.

Page 26 of 31

| | ARGUMENT | COPERNICAN ANSWER |
|---|---|---|
| #72 | *...or at least the refraction must not occur as expected...* | |
| #73 | *...and in particular it must not follow the expected rule for incident and refracted angles...* | |
| #74 | *...and the radius of the earth, the altitude of the refractive air, the amount of refraction of the fixed stars, the refractions and distances of the sun and moon, and so on, indicate that the distance of the fixed stars from earth ought to be far smaller than the Copernican hypothesis requires.* | |
| #75 | *The fixed stars are, in the Copernican hypothesis, so remote that there would be a cessation of refraction of their rays in the lens of a telescope; the telescope would not enlarge them. This disagrees with experience.* | |

| ARGUMENT | COPERNICAN ANSWER |
|---|---|
| Comment: These arguments, regarding refraction of the light from stars, are apparently based on a misunderstanding of the geometry of light rays from a very distant light source, or on the idea that a distant source will mean that a cessation of refraction will occur. Riccioli essentially says that a proper understanding of refraction and geometry answers these arguments. For example, the distance of the stars does not mean the angle rule is violated; in fact "the most subtle calculations [found elsewhere in the *Almagestum Novum*] reveal the opposite".[47] | |
| #76 "The centers of the Earth and the Universe are separated by the radius of the annual orb, so it is uncertain from where we ought to estimate the true altitude of the stars."[48] | "This measure might be estimated from both centers, although by different ways."[49] |
| #77 Admitting the Copernican hypothesis grants license to have any sort of system, arranged around any planet, in the center of the Universe. | Any such system must uphold the celestial phenomena, and none that do are more suitable than that of Copernicus. |

*Acknowledgement*

I thank Christina Graney for her assistance in translating. Not a word was translated without her.

---

47  "Responsum est Negando Maiorem, cuius oppositum initis subtilissime calculis luculentur ostensum est cap. 31. a numero 3. ad 6." (LXXIII Responsum)

48  "Centris terræ et vniuersi per semidiametrum orbis annui seiunctis, incerti essemus, vnde veram altitudinem stellarum æstimare deberemus."

49  "ex vtroque enim æstimanda esset, licet diuerso modo, hæc mensura."

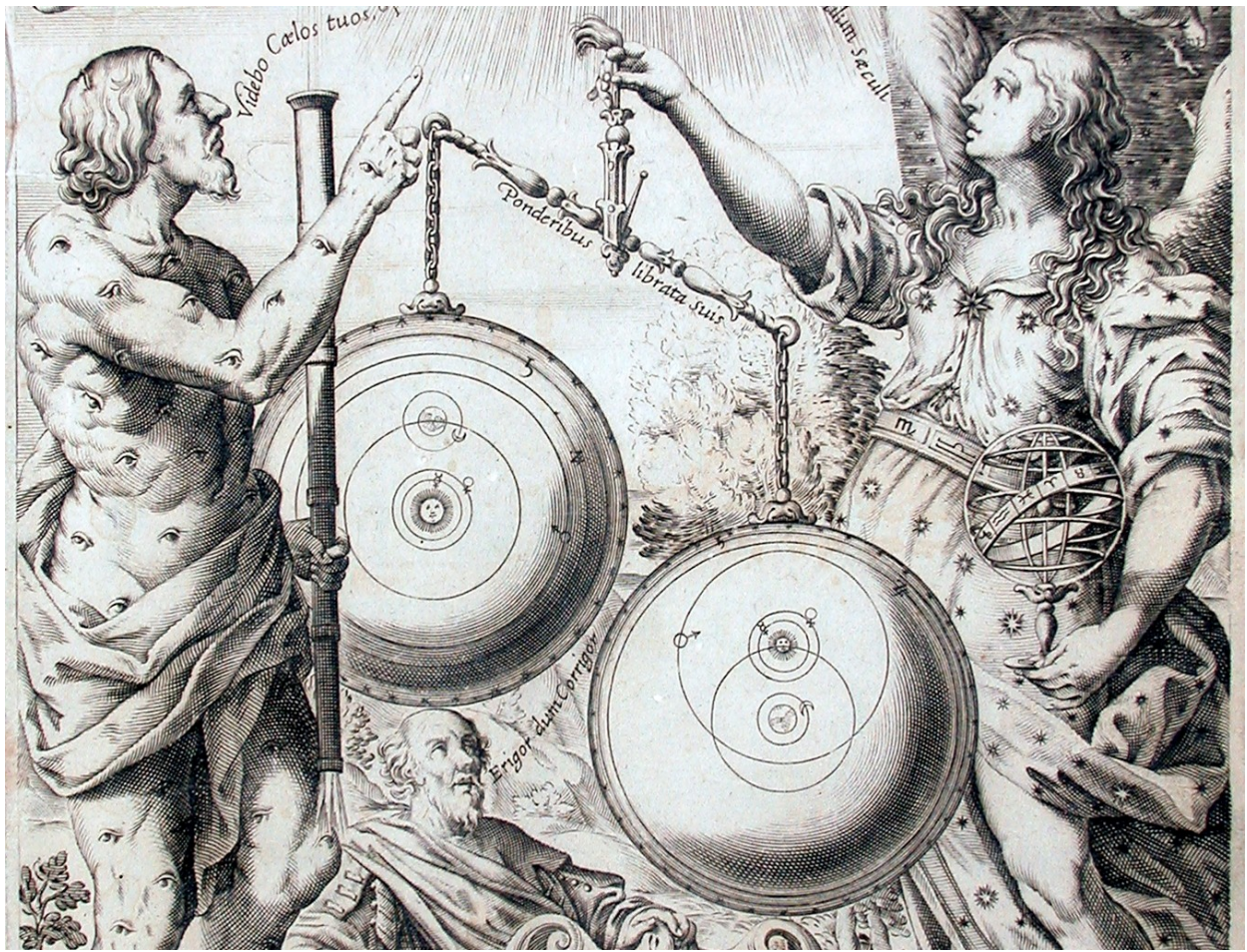

Figure 1: Portion of the frontispiece of Riccioli's 1651 *Almagestum Novum*, showing the heliocentric and geo-heliocentric hypotheses being weighed in a balance by fanciful figures. The geo-heliocentric hypothesis is the weightier of the two.

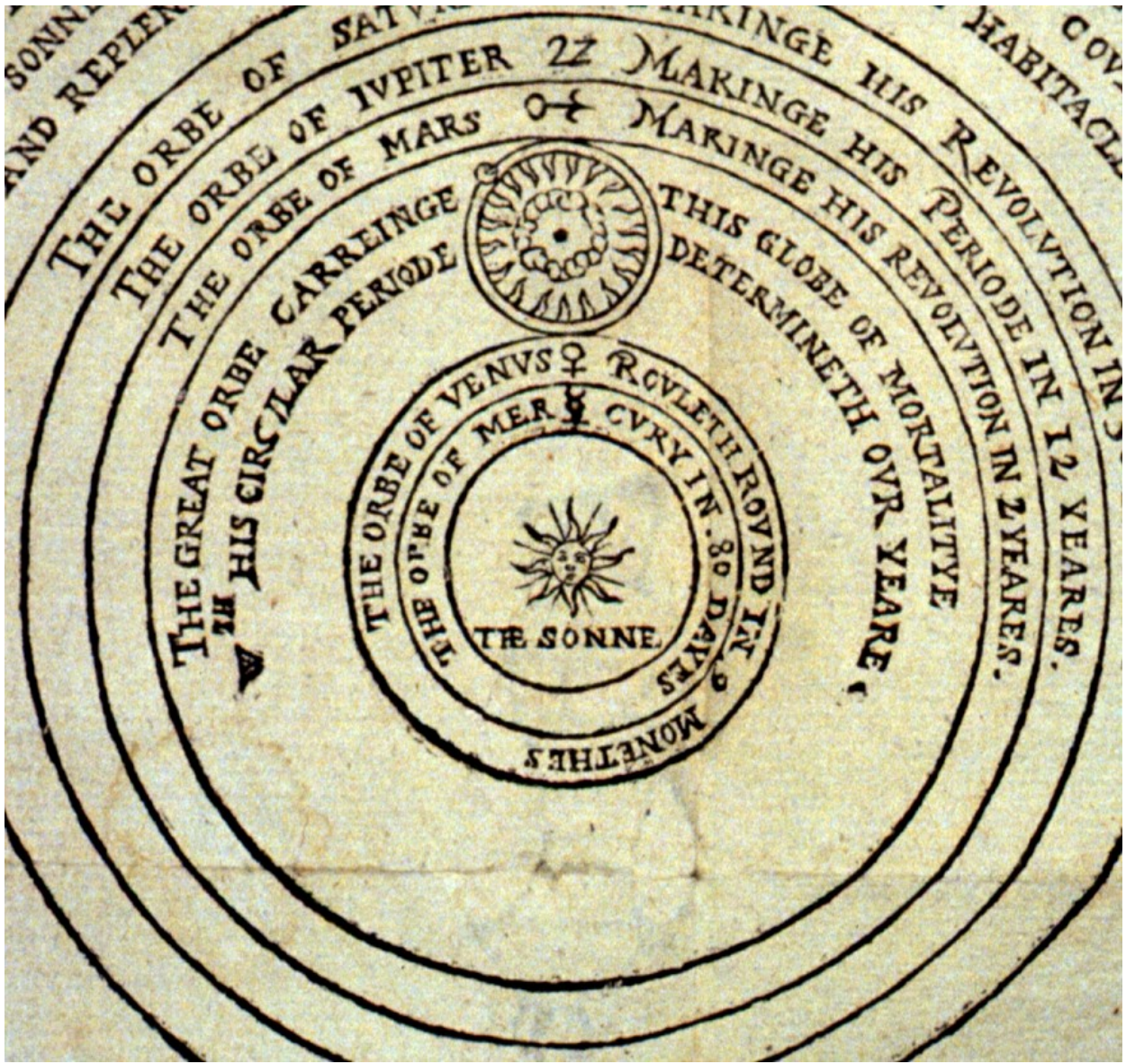

Figure 2: An illustration of the idea that the Earth lies at the center of a spherical elemental system that circles the sun as a whole, and within which the Aristotelian elements and physics is valid (note the elemental fire just within the moon). From Thomas Digges' 1576 *A Perfit Description of the Cælestiall Orbes*.

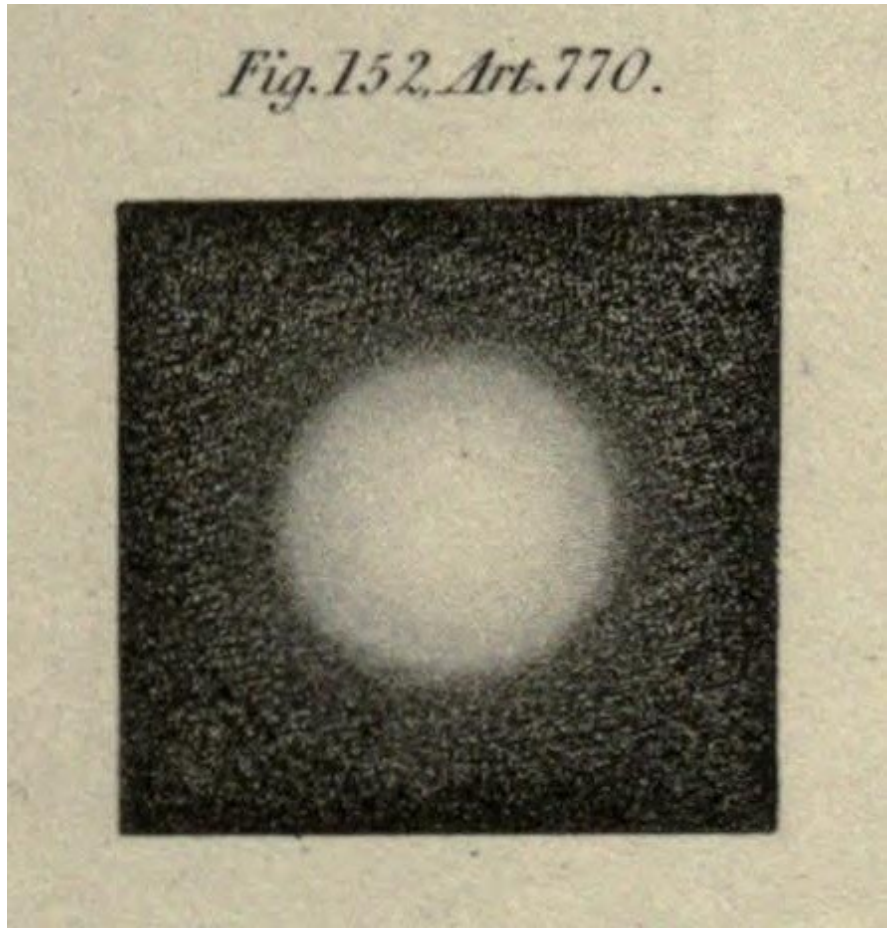

Figure 3: The appearance of a star seen through a small aperture telescope. The star's globe-like appearance is spurious – an artifact of the wave nature of light – but this was not understood by Riccioli and others of the time, who believed they were seeing the physical globes of stars. This illustration is from John Herschel's 1828 *Encyclopædia Metropolitana* article on 'Light' (see Graney and Grayson, "On the Telescopic Disks of Stars...").